# Selfishness as a Virtue in Mobile Ad Hoc Networks

As Related to the 2013 MANIAC Challenge

Isaac Supeene
Department of Electrical and Computer Engineering
University of Alberta
Edmonton, Canada
isupeene@ualberta.ca

Asanga Udugama
Faculty of Electrical Engineering
University of Bremen
Bremen, Germany
adu@comnets.uni-bremen.de

René Steinrücken
Institute of Communication Networks
Technical University of Hamburg-Harburg
Hamburg, Germany
rene.steinruecken@tuhh.de

*Abstract*—Using Mobile Ad Hoc Networks as a way to break the dependency of mobile communications on permanent infrastructure has been an active area of research for several years, but only more recently has consideration been given to the economics of a distributed mobile service. In particular, the topic of handling nodes that act selfishly or maliciously has not yet been explored in detail. In this paper, we discuss the SAVMAN strategy for mobile offloading under the rules of the 2013 MANIAC Challenge.

## I. Introduction

This work developed from the perceived need to introduce a credit-based approach to Mobile Ad Hoc Networks (MANETs) [2]. A credit based system is one in which either a source or destination node must pay some sort of virtual currency to send or receive a packet, which is somehow distributed amongst the intermediate nodes for their service. Such an approach is intended to combat the problem of uncooperative nodes, which can typically be divided into three categories: selfish, malicious, and erroneous/malfunctioning [3]. Selfish nodes participate in the network only to send their own packets, and avoid contributing to the network in order to save their own resources [4]. Malicious nodes, on the other hand, are nodes which actively attempt to disrupt a network, for example by flooding the network with useless packets, or dropping all packets routed through it in a "black hole" attack [1].[1] Malfunctioning nodes are nodes which cannot participate in, or which inadvertently disrupt the operation of a MANET, due to erroneous software or hardware. The observable effects of such nodes are not believed to be qualitatively distinct from those of a malicious node, and we will therefore not consider them further in this paper [2].

The remainder of this paper is structured as follows. In the next section, we discuss how the SAVMAN (Selfishness As a Virtue in Mobile Ad hoc Networks) strategy can maximize throughput under the credit-based system which forms the basis for the 2013 MANIAC Challenge [5]. Section III discusses the testing of our strategy and emulation of the MANIAC Challenge. Section IV provides a retrospective of lessons learned during the challenge itself, and in section V, we present our conclusions.

## II. Maximizing a Node's Balance in the MANIAC Challenge

### A. Victory

The MANIAC Challenge[2] victory conditions are based on two criteria: (a) the maximum balance above 0, and (b) the highest packet delivery ratio [5]. Since in the most realistic scenario, a totally selfish node is concerned only with its own balance, we will focus on criterion (a). (Note that for our purposes, we also do not consider attempting to minimize the balance of the other nodes, since this is only desirable under the artificial constraints of the Challenge itself.)

### B. Maximizing Total Probabilistic Gain

A node performs two main actions to fulfill its role as a forwarding relay: acquiring packets (by bidding on packets auctioned by adjacent nodes) and forwarding packets (by holding its own auctions). The SAVMAN strategy for maximizing a node's balance is based on unifying these two components – in particular, we observe that the optimal strategy for forwarding a packet is inextricably bound to how we acquired the packet. Following this line of reasoning, we derive an equation for *total probabilistic gain* $G_{tot}$ as follows:

---

[1] Other types of maliciousness, such as actively interfering with the wireless communications channel, are not discussed here.

[2] The rules of the 2013 MANIAC Challenge are not given in this text, since the primary audience for this paper is already familiar with them. A full description of the rules can be found in [5].

$$G_{tot} = P_{bid}G_{bid} , \quad (1)$$

where $P_{bid}$ is the probability of winning a bid, and $G_{bid}$ is the *conditional probabilistic gain*, which is the node's prediction of the balance increase it would expect, on average, from packets identical to the one in question, assuming its bid is successful. Our goal is to maximize the total probabilistic gain for each bid.

We further expand $G_{bid}$ as follows:

$$G_{bid} = P_{for}G_{for} + P_{end}G_{end}$$

Here, $G_{for}$ is the *forwarding probabilistic gain*, which is the gain we can expect to achieve from forwarding a packet, and $G_{end}$ is the gain we can expect if we are unable to forward a packet at all. $P_{for}$ and $P_{end}$ are the probabilities associated with each outcome. $G_{end}$ is trivial to calculate based on the rules of the challenge; $P_{for}$ and $G_{for}$ however, are not trivial to calculate directly and independently from one another, so we approximate the term $P_{for}G_{for}$ as $max(P_{fori}G_{fori})$, where the subscript $i$ indicates a value specific to one particular next hop (the one leading to node $i$). Thus, we calculate the probability of success and the most probable gain on forwarding for each available node, and take the maximum of that set:

$$G_{bid} = max(P_{fori}G_{fori}) + P_{end}(B_U - F_U)$$

We split $P_{fori}G_{fori}$ into its constituent cases as follows:

$$P_{fori}G_{fori} = (B_U - B_{Di})P_{succi} + (F_D - F_U)P_{faili} \quad (2)$$

where $P_{succi}$ and $P_{faili}$ are the probabilities of succeeding or failing, respectively, when forwarding to a particular node $i$. $F$ and $B$ indicate the value of the fine and the bid, respectively. Variables subscripted with $U$ are upstream values – that is the values given in the advert on which we're bidding – while variables subscripted with $D$ are the downstream values; $F_D$ is the fine set by us, and is one of the unknowns we are trying to determine, while $B_{Di}$ is the amount bid by the downstream node for our packet, which we must estimate based on that node's previous behaviour.

Given the above equations, we see that our unknown terms are either parameters which we set (such as $B_U$, our bid for the packet, and $F_D$, the fine we set for our own auction), or parameters which we can approximate given specific values for our own budget, bid and fine, and the known history of a previous node. Thus, the entire equation is a function of three variables – the amount we bid on the packet, the fine we set on the packet, and the budget we place on the packet – all of which our strategy has the freedom to choose. In preparation for finding the set of optimal parameters, we will make one more assumption about this function – that it is *unimodal*. That is, we assume that there are no local maxima or minima aside from the single, global maximum. This is admittedly an untested and uncertain assertion, which further research into this area would need to investigate in more detail.

### C. The Fibonacci Cube

Now that the total probabilistic gain is defined in terms of our three unknowns, we employ an extension of the Fibonacci search algorithm [6] in three dimensions. This algorithm operates on effectively the same principle as the one-dimensional Fibonacci search. The search area consists of a rectangular prism, whose bounds are provided at the start, and at each iteration, the largest dimension is chosen to be reduced. The effective value given to a particular plane in the prism is given by the maximum of all points in that plane, which is itself determined by using a two-dimensional Fibonacci Search over the plane. Thus, given plane-wise maxima for each dimension, the dimensions of the prism can be sequentially reduced according to the Fibonacci sequence, eventually providing us with a maximum over the entire volume.

This, of course, requires boundaries for each of the variables – bid, fine and budget. Hard boundaries for bid and fine are specified by the rules of the challenge, and the budget can be no less than 1. In order to provide an upper bound on the budget, we simply define an upper bound equal to the budget of the previous auction, which should cover most reasonable cases.

### D. Avoiding Uncooperative Nodes

It is notable (though not surprising) that $G_{faili}$ cannot be positive. This means that bias towards choosing nodes with a minimal $P_{faili}$ is one of this algorithm's properties. $P_{faili}$ represents our approximation of the probability of success for each possible forwarding node; thus, our selection procedure for packet-forwarding will take into account these probabilities, and tend to avoid uncooperative nodes, once their unwillingness to forward packets has been established.

### E. Predicting the Behaviour of Neighbouring Nodes

As we've seen in the previous sections, there are certain variables in our function which can only be ascertained (or approximated) by attempting to predict the behaviour of other nodes. To this end, we have created three types of *node profile*, which are of the utmost importance in predicting node behaviour. Each node in the network has these three profile types associated with it, and each profile assists in predicting that nodes behaviour in one particular aspect of the challenge.

The first type of node profile is the Bidding Profile. This profile assists in determining probable bids we can expect from other nodes given a particular auction, thus enabling us to make competent decisions both as bidders and auctioneers. This bidding profile is updated by observing other nodes' responses to auctions and correlating that information with the current topology and the parameters of the auction.

The second type of profile is the Auction Profile. This is used to determine which node is likely to be chosen as a winner for a particular auction – for example, a node might always choose the cheapest bid, or the cheapest bid that has a chance of getting to the destination within the budget, or simply a random bid.

The third, and most important profile type is the Reliability profile. This profile observes the success rate of a particular node under various topological circumstances, and uses this information to select an appropriate next node for forwarding.

Through these three profiles, the SAVMAN strategy is able to predict the behaviour of other nodes in order to optimize its own behaviour and profit.

For further details about the strategy, the reader is invited to view our implementation, which is stored in the maniacchallenge/2013 github repository [9].

## III. Testing and Simulations

### A. The Two Testing Approaches

Validating and improving the theoretical approach before the MANIAC Challenge was crucial to avoid unforeseen obstacles. Two approaches were considered during the preparation for the challenge. The first approach consisted of a single-machine simulation, whereas the second approach consisted of a testbed with netbooks functioning as nodes in the network.

While both approaches require multiple changes and adaptations of the MANIAC API, the requirement were set to minimize the impact of these changes in comparison to the final challenge.

The simulation software, implemented in Java, fulfills this requirement by allowing the assignment of virtual devices to different (local) IP addresses on one machine. This approach leaves the packet handling of the API untouched and comparable to the final challenge. The simulation then creates virtual backbones which inject packets into the network.

The Nexus 7 tablets in the challenge were to be provided with network information through the olsrd - txtinfo plugin [7]. To emulate this behavior, a software class provides each virtual node with a fake network topology. This topology can be set up to appear as a multi-hop network and allows each virtual node to behave differently according to its own strategy. This software approach also allows the simulation of lost packets similar to a real network.

Tests with the simulation provided satisfying results with smaller networks. Due to timing constrains in the final MANIAC Challenge and the increased processing power and computation time which the single-machine simulation required for larger networks, the second approach provided a test environment more comparable to the final challenge.

The second approach emulates Nexus 7 behavior in an OLSR network. The open-source project olsrd [8] allows the creation of an optimized link state network easily, and is the basis of this approach. Machines in the network run olsrd and the txtinfo plugin, as well as an adapted version of the MANIAC API. Selected nodes act as backbones and provide packets for the nodes to deliver. This approach includes realistic timeouts, packet collision and network changes. After completing the setup, the testbed revealed the distinct advantages and disadvantages of different strategies and node behavior, and enabled suitable preparation for the challenge.

### B. Test-bed Considerations

The Linux based OLSR network of the test-bed consisted of 6 netbooks and 2 Nexus 7 devices, connected over ad-hoc mode WLAN. 2 of the netbooks were configured to run an emulated version of the backbone, while the rest of the netbooks were deployed with the adapted version of the MANIAC API as competing MANIAC nodes. The Nexus 7 devices were deployed with the SAVMAN strategy. In this section, we describe some of the important aspects considered when building the test-bed.

**Multi-hop connectivity**: Wireless proximity of the nodes in the test-bed resulted in single hope OLSR connectivity between all the nodes. To test our strategies in a fairly realistic manner, we setup Media Access Control (MAC) layer filtering on the nodes to form a multi-hop OLSR network based on the different network topologies that we identified. The IPTABLES facility in Linux was used to setup MAC filtering.

**Topology**: We identified a number of scenarios in which to test the SAVMAN strategy. Each of these scenarios was tested with a particular multi-hop network topology.

**Network Monitoring:** One of the aims in testing the SAVMAN strategy in the test-bed was to determine how well it worked against the other nodes in the OLSR network. To identify the performance of each node in the OLSR network, we developed a packet sniffer application that dumped statistics on all auctions, bids, bid wins and the balances of each of the nodes. This application used the PCAP facility in Linux to capture all packets in the OLSR network.

## IV. Lessons Learned

One of the crucial lessons learned in the MANIAC Challenge was related to the change in network behaviour due to heterogeneous strategies. While any given strategy may have worked well in a controlled environment, the entire collection of strategies that were employed did not mix well together. For example, certain strategies would attempt to repeatedly undercut one another in the bidding process, leading to extremely low bidding prices and subsequent auction budgets, which could lead to adjacent nodes simply not bothering to forward the packet.

A related point of interest was a modification made by the Brazilian team to their strategy – their observation was that certain teams were putting out auctions with a budget of 1, and their nodes were then winning these packets and forwarding them directly to the destination. Of course the selfish behaviour of setting the auction budget to 1 is harmful to the network in general, despite being a potentially effective strategy. Therefore the Brazilian team chose to make a modification to their strategy in subsequent rounds which would attempt to acquire and drop all packets auctioned with a sufficently low budget. They found that after implementing this, nodes which had been bidding on packets at a reasonable price, then auctioning them off at a very low price actually adapted their behaviour and ceased giving them packets for a price of 1. However, this only increased the problem of cost-undercutting mentioned above, by creating a black hole for all packets acquired by low-bidding nodes.

## V. Conclusion

Mobile Ad Hoc Networks (MANETS) rely strongly on cooperation between nodes due to their distributed nature, which naturally makes them highly vulnerable to the effects of selfish and malicious nodes. This can be controlled by a credit-based system, but only to the extent that the system itself does not open further opportunities for exploitation.

In this paper, we have seen how the MANIAC framework permits a node to act entirely in its own interest, while still contributing positively to the package delivery rate of the MANET. We have also described the SAVMAN strategy for mobile offloading, which is based on maximizing a mathematical function whose parameters are determined by carefully observing the behaviour of other nodes, and we have further described how this strategy was tested by creating a simulation of the MANIAC Challenge and pitting our strategy agains some artificial opponents. Lastly, we have provided a retrospective of the challenge itself, containing observations and lessons learned which we hope will be useful for future research in this area.

References

[1] S. Djahel, F. Naït-abdesselam, and Z. Zhang, "Mitigating packet dropping problem in mobile ad hoc networks: proposals and challenges," IEEE Communications Surveys & Tutorials, vol. 13, no. 4, fourth quarter 2011, pp. 658–672

[2] S. Zhong, J. Chen, and Y. R. Yang, "Sprite: a simple, cheat-proof credit-based system for mobile ad hoc networks," IEEE INFOCOM 2003

[3] M. Schütte, "Detecting selfish and malicious nodes in MANETs," Seminar: Sicherheit in Selbstorganisierenden Netzen, HPI/Universität Potsdam, Sommersemester 2006.

[4] K. Balakrishnan, J. Deng, and P. K. Varshney, "TWOACK: preventing selfishness in mobile ad hoc networks," Wireless Communications and Networking Conference, 2005 IEEE (Volume: 4).

[5] MANIAC 2013 Challenge, http://2013.maniacchallenge.org/

[6] http://en.wikipedia.org/wiki/Fibonacci_search_technique

[7] OLSRd txtinfo plugin, http://www.olsr.org/?q=txtinfo_plugin

[8] OLSR deamon, http://www.olsr.org

[9] https://github.com/maniacchallenge/2013/tree/master/strategies%20(implementations)/SAVMAN